\documentclass[twocolumn,showpacs,aps,prl,amsmath,amssymb,nofootinbib,nobibnotes,floatfix]{revtex4-1}

\usepackage{hyperref}
\usepackage{amsfonts}
\usepackage{amsmath}
\usepackage{mathrsfs}
\usepackage{epsfig}
\usepackage{graphicx}               
\usepackage{url}
\usepackage{hyperref}
\usepackage{float}
\usepackage{color}
\usepackage[utf8]{inputenc}

\usepackage{graphicx}
\usepackage{epsf}

\newcommand{\ssec}[1]{{{\em #1.}---}}
\usepackage{comment}

\newcommand{\nc}{\newcommand}

\nc{\beq}{\begin{equation}}
\nc{\eeq}{\end{equation}}
\nc{\beqa}{\begin{eqnarray}}
\nc{\eeqa}{\end{eqnarray}}

\usepackage{slashed}

\newcommand{\lsim}{\!\mathrel{\hbox{\rlap{\lower.55ex \hbox{$\sim$}} \kern-.34em \raise.4ex \hbox{$<$}}}}
\newcommand{\gsim}{\!\mathrel{\hbox{\rlap{\lower.55ex \hbox{$\sim$}} \kern-.34em \raise.4ex \hbox{$>$}}}}

\def\be{\begin{equation}}
\def\ee{\end{equation}}

\newcommand\affspc{\vspace{4pt}}

\usepackage{footmisc}
\usepackage{setspace}
\begin{document}

\title{Cosmic Censorship Upheld in Spheroidal Collapse of Collisionless Matter}

\author{William E.\ East}
\affiliation{Perimeter Institute for Theoretical Physics, Waterloo, Ontario N2L 2Y5, Canada \affspc}

\begin{abstract}
We study the collapse of spheroidal configurations of collisionless particles in
full general relativity. This setup was originally considered by Shapiro and
Teukolsky (1991), where it was found that prolate configurations with
a sufficiently large semimajor axis gave rise to diverging curvature, but no
apparent horizon. This was taken as evidence for the formation of a naked
singularity, in violation of cosmic censorship.  We revisit such
configurations using different coordinates/slicing, and considering a range of
values for the semimajor axis and eccentricity of the initial matter
distribution, and find that the final state in all cases studied is a black
hole plus gravitational radiation.  Though initially distorted, the proper
circumferences of the apparent horizons that are found do not significantly
exceed the hoop conjecture bound.  Configurations with a larger semimajor
axis can produce strong gravitational radiation, with luminosities up to $P_{\rm
GW}\sim 2\times 10^{-3}c^5/G$. 
\end{abstract}

\maketitle

\ssec{Introduction}%
Unhalted, gravitational collapse in Einstein's theory of
general relativity will lead to the formation of singularities
where the theory breaks down.
Remarkably, it is conjectured that if one excludes regions
hidden from far away observers by black hole horizons, 
Einstein's theory generically
retains its ability to predict the evolution of spacetime. 
This behavior is referred to as cosmic censorship~\cite{Penrose:1969pc}.
(Here, we refer specifically to the \emph{weak} cosmic censorship conjecture.)
Violations of cosmic censorship have been shown to occur in the fine-tuned
configurations of critical collapse~\cite{Choptuik:1992jv}, as well as in spacetimes
with dimension higher than four~\cite{Lehner:2010pn,Figueras:2015hkb,Figueras:2017zwa}.
However, for more generic initial data in 3+1 dimensions, cosmic censorship
has shown to hold in the numerous cases where it has been studied~\cite{Wald:1997wa}.

One possible exception to cosmic censorship is the work of Shapiro and
Teukolsky~\cite{Shapiro:1991zza} following the collapse of prolate
configurations of collisionless particles. In that work, one 
configuration was found to exhibit a blowup in the spacetime curvature
in two spindle regions lying \emph{outside} the collapsing matter, an indication 
this was not due to just a shell crossing singularity in the matter.
\footnote{A formal statement of the cosmic censorship conjecture requires 
some ``suitability" condition be placed on the matter fields, which could
exclude singularities such as fluid shocks or matter shell crossings
that could occur even without gravity~\cite{Wald:1997wa}. Here, following~\cite{Shapiro:1991zza}, we 
consider collisionless matter, which can develop caustics, but do not
consider such caustics as violations of cosmic censorship.}
No apparent horizon was found before the simulation became unreliable
and could no longer be evolved, and this, along
with the lack of turn around of null geodesics in the vicinity of
the blowup, was taken as evidence that this was a naked singularity, 
in violation of cosmic censorship.
However, since the calculation could
not be continued past the blowup, 
the possibility of the spacetime containing 
an event horizon that would hide the
final stages of gravitational collapse was not conclusively ruled out. 
In particular, the absence of an apparent horizon in some gauge does not 
rule out an event horizon. In fact, one
can slice a Schwarzschild spacetime in a
way that approaches the singularity without
containing outer trapped surfaces~\cite{Wald:1991zz}.  

The study of such prolate configurations as candidates for
violating cosmic censorship was motivated by Thorne's hoop
conjecture~\cite{thorne_hoop}.  
Though lacking a precise formulation, the hoop
conjecture roughly states that a black hole will form if and only if some mass
$M$ can be localized in a region whose circumference in every direction
satisfies $\mathcal{C}\lesssim 4 \pi M$. (Here and throughout we use geometric
units with $G=c=1$.)
For example, the collapse of an infinite cylindrical distribution of matter
will not a form a black hole (but will form a singularity)~\cite{thorne_hoop}. 
The hoop conjecture can be violated in the presence of negative energy, for
example, by cylindrical black holes in anti-de Sitter spacetimes~\cite{Lemos:1994xp,Lemos:1997bd}, or due to a scalar field with 
negative potential~\cite{East:2016anr}, where arbitrarily elongated black holes
can be formed, but seems to be robust otherwise. 
Hence the motivation to study the collapse of very prolate distributions
of matter which (at least initially) lie outside the hoop conjecture bound
to form a black hole~\cite{Nakamura:1988zq}.
(See Ref.~\cite{Andreasson:2012hx} for an analytic proof of black hole
formation in a spherical symmetric collisionless matter configuration.)

Despite follow-up work by numerous authors, including relaxing the requirement
of axisymmetry and using higher resolution~\cite{Shibata:1999va,Yoo:2016kzu},
utilizing excision to the causal future of the curvature blowup~\cite{Okounkova:gr21},
and extending to 5 dimensional spacetimes~\cite{Yamada:2011br}, the question of
whether the configurations studied in Ref.~\cite{Shapiro:1991zza} violate cosmic
censorship has remained unanswered.
Here, we revisit the problem, using somewhat different methods that allow us 
to rescue cosmic censorship and determine the ultimate fate of such spacetimes. 
We show that the final state is in fact a black hole with, in some cases significant, gravitational radiation.

\ssec{Methodology}%
We consider the same family of initial conditions as in Ref.~\cite{Shapiro:1991zza}, consisting of a prolate spheroidal
distribution of collisionless matter, initially at rest,
that is axisymmetric and has no angular momentum. This family is 
parameterized by a semimajor axis length $b$ (in units of the total mass $M$), 
and eccentricity $e=\sqrt{1-a^2/b^2}$
(where $a$ is in the equatorial radius). 
In the Newtonian limit, the spheriods have homogeneous density.
At $t=0$, the spatial metric is conformally flat $\gamma_{ij}=\Psi^4 \delta_{ij}$
and the extrinsic curvature is zero $K_{ij}=0$. 
See Refs.~\cite{Nakamura:1988zq,Yoo:2016kzu} for further details on the initial data.

In this work, we focus on very prolate cases. In Ref.~\cite{Shapiro:1991zza}, the cases
considered were $e=0.9$ with $b/M=2$ (prompt collapse to a black hole)
and $b/M=10$ (candidate for cosmic censorship violation). Here we consider
a number of cases with $e=0.9$ and $2 \leq b/M \leq 20$. 
We also consider
one case with larger eccentricity, namely $b/M=10$ with $e=0.95$.

We evolve the Einstein-Vlasov equations describing 
a distribution of collisionless matter coupled to gravity using the methods
of Ref.~\cite{Pretorius:2018lfb} for evolving massive particles.
For gauge conditions at $t=0$, we choose the lapse to be $\alpha=\Psi^{-4}$
and the shift to be zero $\beta^i=0$. However, we carry out the initial part of
the evolution in harmonic gauge. Around the time of collapse, we transition to
a damped harmonic gauge~\cite{Lindblom:2009tu,Choptuik:2009ww} (specifically
the $p=1/4$ version used in Ref.~\cite{East:2012mb}), which we find helps control the
strong oscillations in the coordinate shape of the black hole as it settles
down.
In contrast, in Ref.~\cite{Shapiro:1991zza}, maximal slicing and isotropic spatial coordinates
were used.

We search for apparent horizons---outermost marginally outer trapped surfaces---using a flow method~\cite{Pretorius:2004jg}.
Once found, we track the evolution of the horizon, measuring several properties including its
area---from which a mass $M_{\rm BH}$ can be calculated---and its proper circumferences in the polar and equatorial directions,
$\mathcal{C}_{\rm p}$ and $\mathcal{C}_{\rm eq}$.
In the following, we use black hole to refer to the apparent horizon, though
for all cases we track the apparent horizon to sufficiently late times that it
should become a good approximation for a time slice of the event horizon. 
The gravitational radiation is measured by calculating the
Newman-Penrose scalar $\psi_4$.  We also compute the Kretschmann scalar,
obtained from contracting the Riemann tensor with itself $I=R^{abcd}R_{abcd}$,
as well as the matter density, computed from the stress-energy tensor
$\rho=-T^a_a$.

We restrict to axisymmetry, which allows us to use a computational domain with two spatial
dimensions.
Most results presented below are obtained using $N=1.6\times10^6$ particles and
an adaptive mesh refinement simulation grid where the finest resolution
is $dx\approx0.02M$ (for $b/M\leq 12$ and $e\leq 0.9$) or $dx\approx0.01M$ (otherwise).
For select cases, we also perform resolution studies to establish
convergence using $0.75\times$ and $1.5\times$
the grid resolution, and $0.75^4\times$ and $1.5^4\times$ as many particles.
Details on numerical convergence can be found in the appendix.

\begin{figure}
\begin{center}
\includegraphics[width=\columnwidth,draft=false]{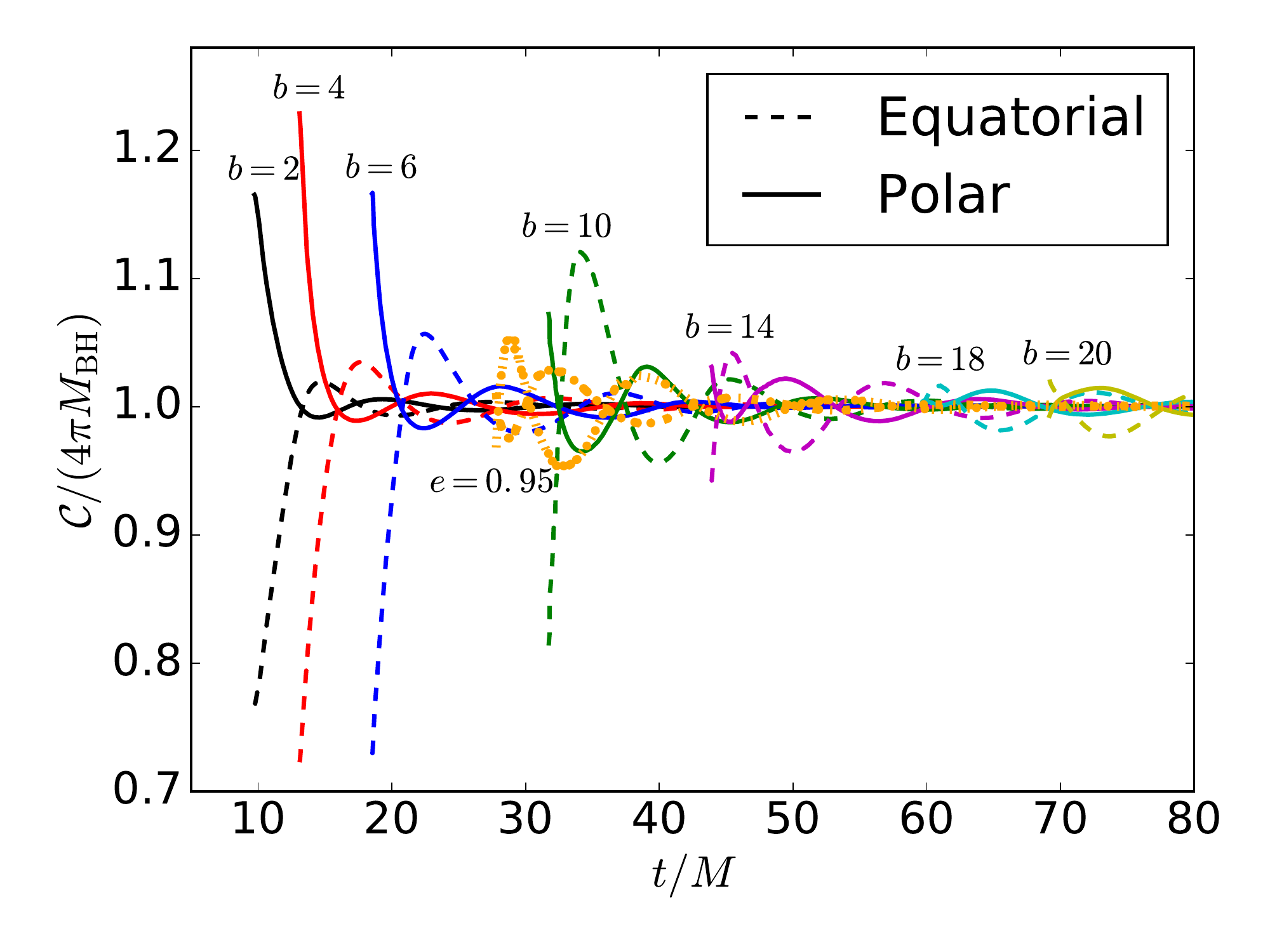}
\end{center}
\caption{
	The polar (solid lines) and equatorial (dashed lines)
	proper circumferences of the apparent horizons found
	for various values of the semimajor axis length $b$ (in
	units of the total mass). The circumference is normalized
	by $4\pi M_{\rm BH}$, where $M_{\rm BH}$ is the mass of 
	the apparent horizon, to indicate the relation to the 
	hoop conjecture bound.
    All curves are for initial data with $e=0.9$, except for the dotted (orange)
    curves which correspond to $e=0.95$ and $b=10$.
\label{fig:ah_circ}
}
\end{figure}

\ssec{Results}
Our main result is that we are able to evolve all cases
considered here until they settle towards a final state, 
which we find to be a black hole containing all the matter,
along with gravitational radiation. 
Configurations with smaller values of
$b$ (or larger values of eccentricity, in the case with $e=0.95$) 
form black holes more promptly, while those
with larger values take longer to collapse---both perpendicular
to the symmetry axis, and along the symmetry axis.
In Fig.~\ref{fig:ah_circ}, we
show the polar and equatorial circumferences of the apparent horizon, beginning
when one is first found, for a number of cases.
The matter configuration has collapsed sufficiently
that the values are not more than $\sim 25\%$ above $4\pi M_{\rm BH}$, 
and thus not in serious violation of the approximate inequality of the hoop conjecture.
(Though we note that, since our horizon finding algorithm
relies on having a sufficiently good guess for the shape,
we cannot exclude the existence of a more distorted horizon 
at earlier times.)
After formation, the horizons then exhibit damped oscillations between 
being prolate and oblate as they ring down towards a stationary state.

\begin{figure}
\begin{center}
\includegraphics[width=\columnwidth,draft=false]{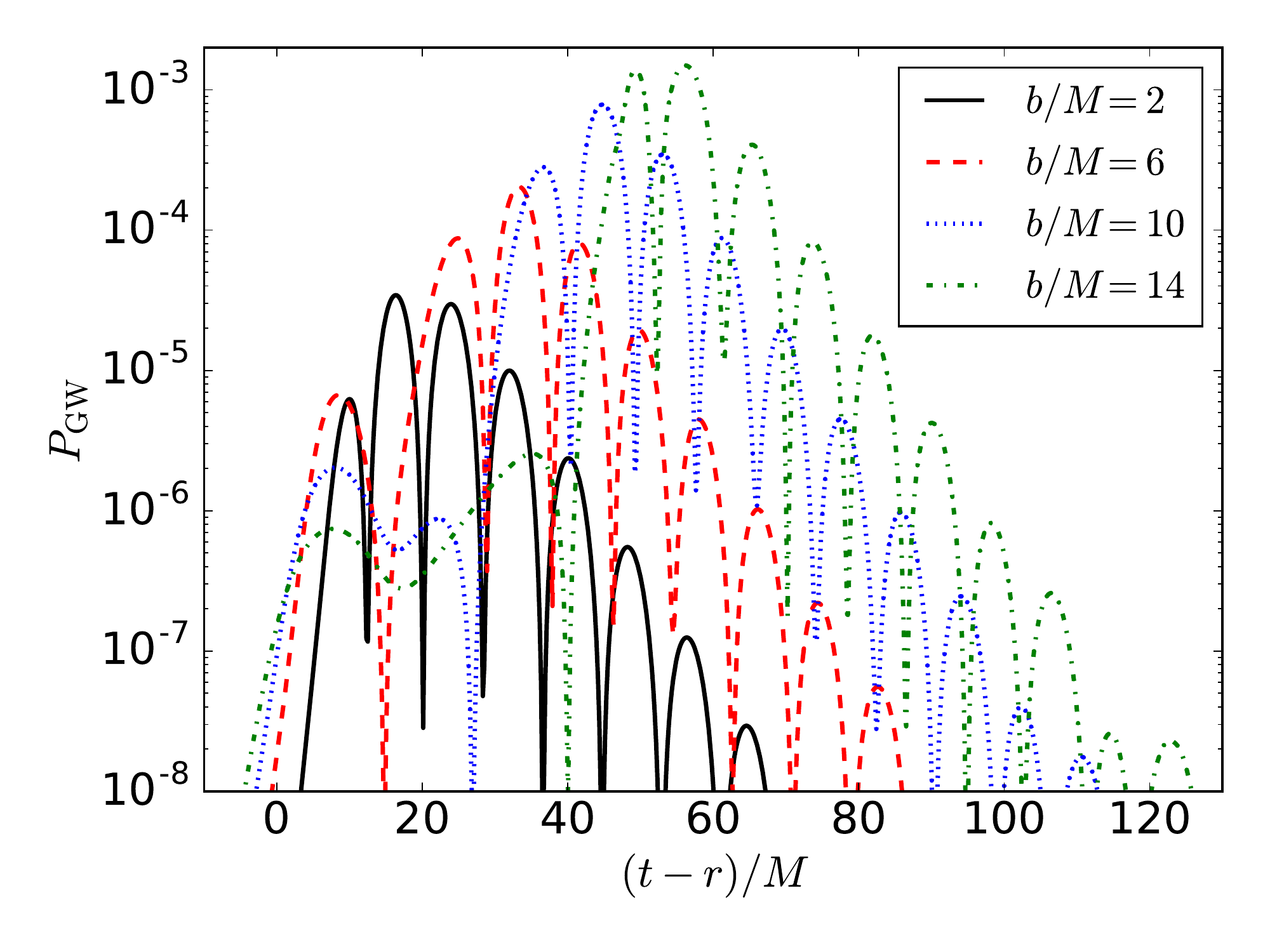}
\end{center}
\caption{
	The gravitational wave power as a function of time
	for cases with various values of the semimajor axis and $e=0.9$.
\label{fig:pgw}
}
\end{figure}

All the matter ends up in the black hole for every case 
studied here. However, a non-negligible amount of energy is radiated
away in gravitational radiation. In Fig.~\ref{fig:pgw},
we show the gravitational wave power for select cases.
The power peaks around the time of black hole formation---reaching 
as high as $P_{\rm GW}\sim0.002$ in some cases---and then dies away exponentially, again showing the characteristic
quasinormal mode ringing. 

\begin{figure}
\begin{center}
\includegraphics[width=\columnwidth,draft=false]{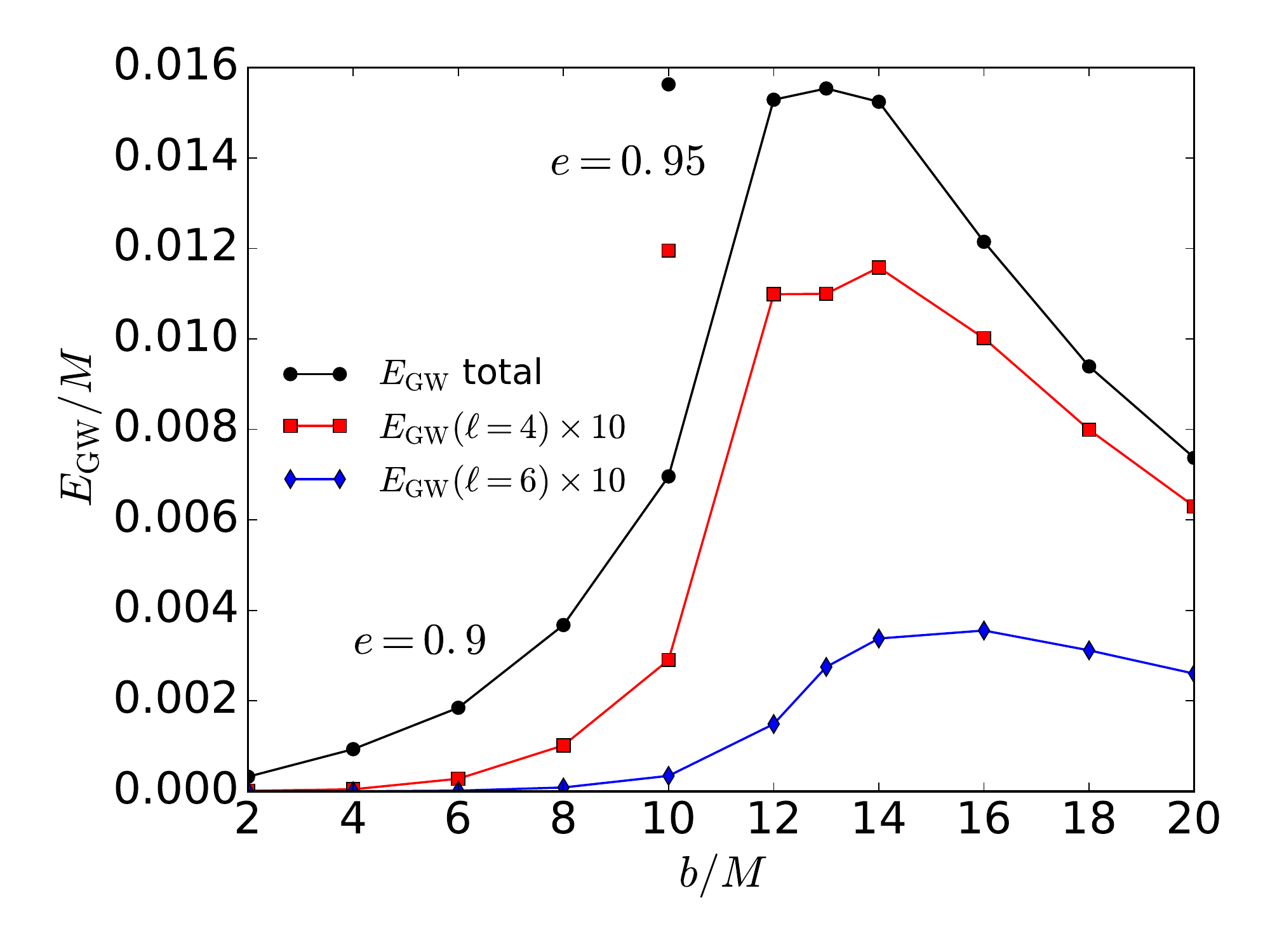}
\end{center}
\caption{
	The total energy emitted in gravitational waves as
	a function of the semimajor axis $b$. The energy is predominately
	due to an $\ell=2$ component, but we also show the amount
	in the $\ell=4$ and $\ell=6$ components, scaled up by
	a factor of $10$ to be visible on the graph.
    The points connected by lines correspond to $e=0.9$, while the
    two unconnected points above correspond to $e=0.95$.
\label{fig:egw}
}
\end{figure}

We also show the
total energy in gravitational radiation as function
of the semimajor axis length in Fig.~\ref{fig:egw}. 
Cases with smaller values of $b/M$ are already close to being black holes at the initial
time and do not emit significant radiation.
For $e=0.9$, this is maximized at $b/M\sim12$--$14$, with $E_{\rm GW}\sim0.015M$.
A similar amount of energy is radiated for $b=10$ and $e=0.95$.
The difference from the total mass $M-E_{\rm GW}$ matches the measured mass of the black hole
at late times to better than $0.2\%$ for all cases.
The amount of gravitational radiation is significant for an axisymmetric
spacetime. For comparison, an equal mass head-on collision of two black 
holes falling from rest releases $0.06\%$ in gravitational radiation~\cite{Sperhake:2011ik}, 
while an ultrarelativistic collision
releases $15\%$ of the total mass in gravitational waves, and has a
peak luminosity of $P_{\rm GW}\sim0.01$~\cite{East:2012mb,Sperhake:2008ga}.

Most of the gravitational wave energy is due to the $\ell=2$ angular component,
but in Fig.~\ref{fig:egw}, we also show the subdominant contributions from the
$\ell=4$ and 6 components.
(The odd $\ell$ components are suppressed by the symmetry of the initial data, 
though we do not explicitly enforce the equatorial symmetry in the placement of particles,
nor during evolution.)

\begin{figure}
\begin{center}
\includegraphics[width=\columnwidth,draft=false]{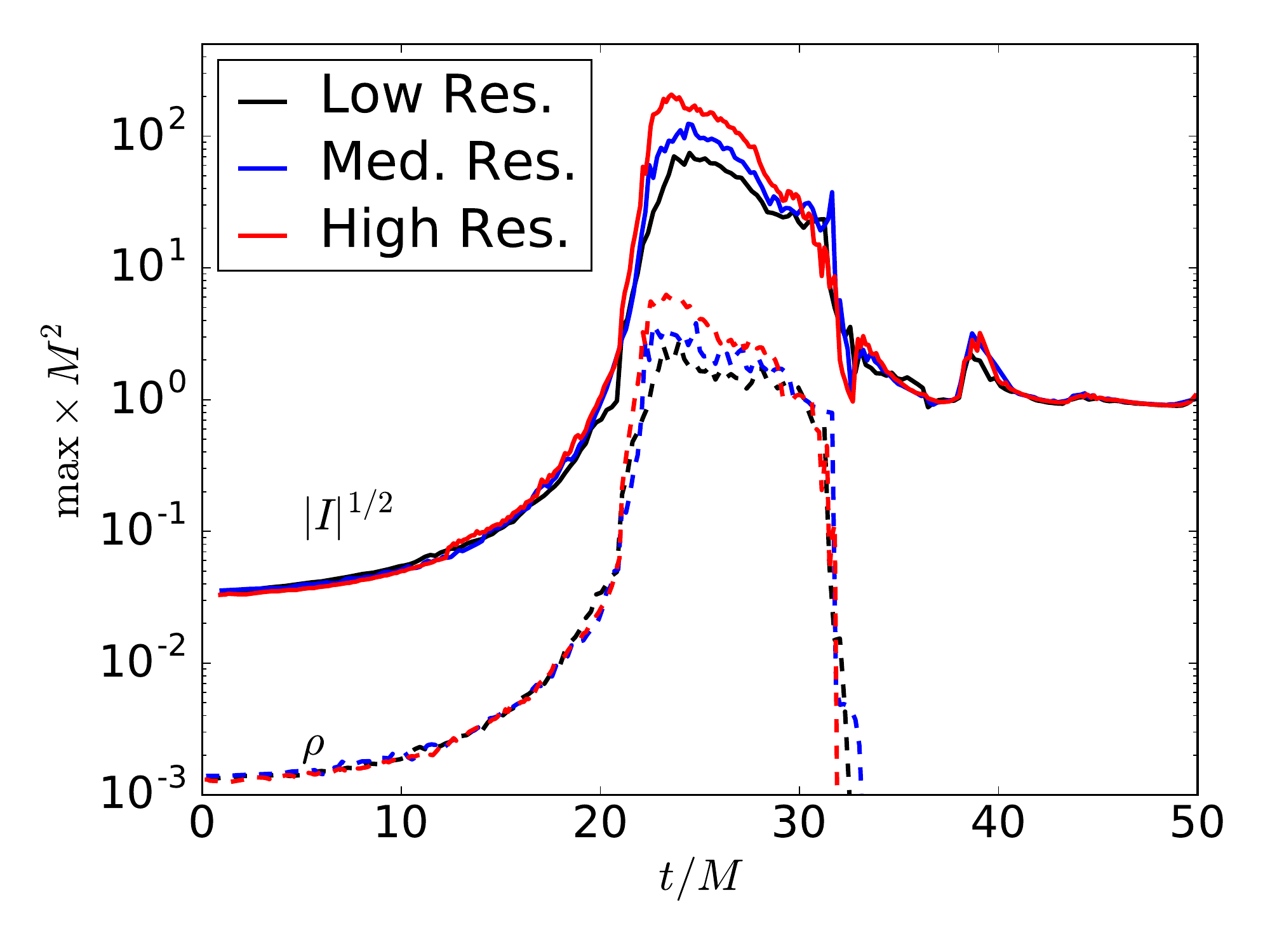}
\end{center}
\caption{
	The maximum value of the particle density (dashed lines) and $|I|^{1/2}$ (solid lines)
	as a function of time (both in units of $1/M^2$), for $b/M=10$ and
	$e=0.9$ and three different resolutions.
	At around $t/M=30$ an apparent horizon is found, the interior of 
	which is excluded from this calculation. 
\label{fig:curv}
}
\end{figure}

Focusing on the $b/M=10$, $e=0.9$ case considered in Ref.~\cite{Shapiro:1991zza},  we
also find a blowup in the curvature around the same time. As shown in
Fig.~\ref{fig:curv}, the maximum value obtained at blowup increases with
resolution. However, in contrast to Ref.~\cite{Shapiro:1991zza}, we find that the
maximum in $I$ always occur in a region where the matter density is nonzero
(this was also found in Ref.~\cite{Yoo:2016kzu}), and in fact tracks the blowup in
density, as shown in Fig.~\ref{fig:curv}. This indicates that this is just due to
the development of caustics in the matter (see, e.g., Ref.~\cite{Giblin:2018ndw}).  

\begin{figure}
\begin{center}
\includegraphics[width=\columnwidth,draft=false]{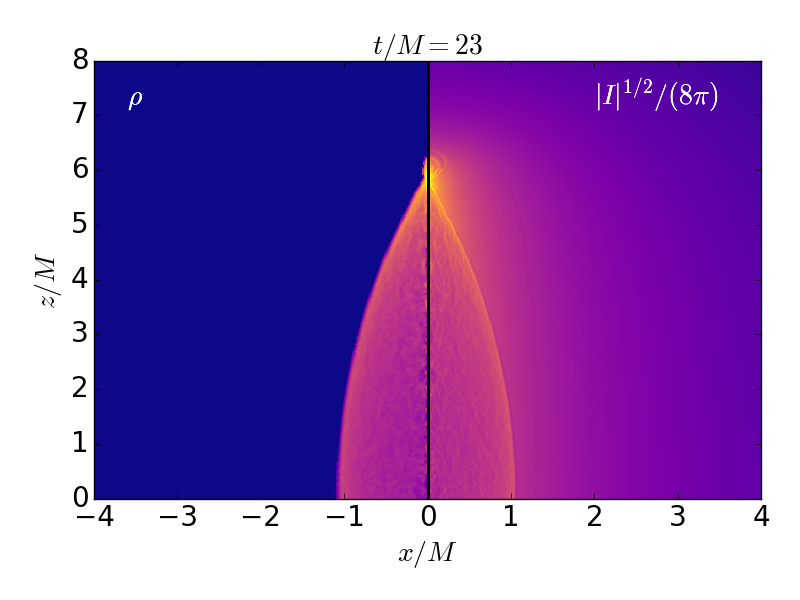}
\includegraphics[width=\columnwidth,draft=false]{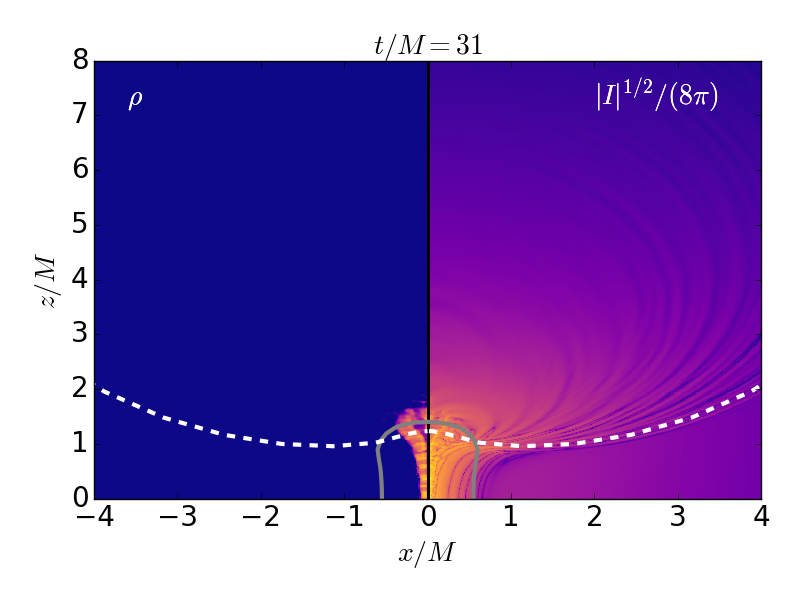}
\end{center}
\caption{
    Snapshots of the particle density (left half) and $|I|^{1/2}/(8\pi)$ (right half), 
    at the approximate times when a caustic first forms (top) and when an apparent
    horizon is first found (indicated by the gray curve; bottom) for $b/M=10$ and $e=0.9$.
    The color scale is logarithmic from $2\times 10^{-4}$ to $2$ in units of $1/M^{2}$.
    The dashed white curve in the bottom panel
    represents the position of a set of null geodesics spreading out in all 
    directions from the point of maximum density and curvature in the top panel.
\label{fig:rho}
}
\end{figure}

This is further illustrated in Fig.~\ref{fig:rho} where 
we show snapshots of both the matter density and curvature. 
By integrating null geodesics outward from the caustic to the wave zone, 
we have explicitly checked that the caustic is ``visible" at null infinity.
We are able to continue the calculation past
the formation of the caustic (which is mild, e.g., compared to the singularity
associated with trapped surfaces). 
The matter continues rapidly collapsing, and a short time later an apparent horizon is
found which envelops all the matter and the region of higher curvature. The curvature
outside quickly approaches the value of an isolated black hole.

\ssec{Conclusion}%
We have followed the relativistic collapse of very prolate spheroidal
configurations of matter, revisiting a scenario originally studied
in Ref.~\cite{Shapiro:1991zza}, and put forth as evidence against cosmic censorship.
With our different choice of slicing and coordinates, we do not find a blowup of curvature
peaked outside the matter region as in Ref.~\cite{Shapiro:1991zza}, and we are able
to follow the evolution through to the asymptotic end state. We see 
that a black hole does form, swallowing the matter, and censoring the interior
singularity.

The original motivation for investigating this scenario for possible violations
of cosmic censorship was that it seemingly pitted the hoop conjecture, which
dictates that sufficiently elongated matter configurations should not form
black hole horizons, against the generic tendency of unhalted relativistic
collapse to form singularities.  We find here that the spacetime dynamics
unfolds in such a way that, even in the case of collisionless particles, the
matter collapses to where it can be surrounded by a horizon that is not too
elongated.  Thus, neither the hoop conjecture, nor cosmic censorship appear to
be violated.  This rapid and violent collapse does, however, leave its imprint
in the strong gravitational radiation, which for some cases is comparable to a
quasicircular binary black hole merger in peak luminosity and the fraction of
the total mass of the spacetime.

\ssec{Acknowledgements}%
I thank Maria Okounkova and Frans Pretorius for stimulating discussions. 
This research was supported in part by Perimeter Institute for Theoretical
Physics. Research at Perimeter Institute is supported by the Government of
Canada through the Department of Innovation, Science and Economic Development
Canada and by the Province of Ontario through the Ministry of Economic
Development, Job Creation and Trade.  Computational resources were provided by
the Symmetry cluster at Perimeter Institute and the Perseus cluster at
Princeton University.

\bibliographystyle{apsrev4-1.bst}
\bibliography{ref}

\appendix
\section{Numerical convergence}
We study the collapse of spheroidal distributions of collisionless particles
using the code described in~\cite{Pretorius:2018lfb}. The massive particles
follow geodesics on the evolving spacetime, and contribute to the stress-energy
tensor used to evolve Einstein's equations.  Though the metric and geodesics are
evolved using fourth order accurate methods, the stress-energy calculation is
only second order accurate, which sets the overall accuracy of the solution.
We use adaptive mesh refinement, with the grid structures determined by
truncation error estimates. When performing a convergence study, we adjust the
truncation error threshold to be consistent with the expected second order
convergence in the grid resolution and (as described in~\cite{Pretorius:2018lfb}) 
increase the number of particles by the fourth power of the increase in
the linear resolution.  

\begin{figure}[H]
\begin{center}
\includegraphics[width=\columnwidth,draft=false]{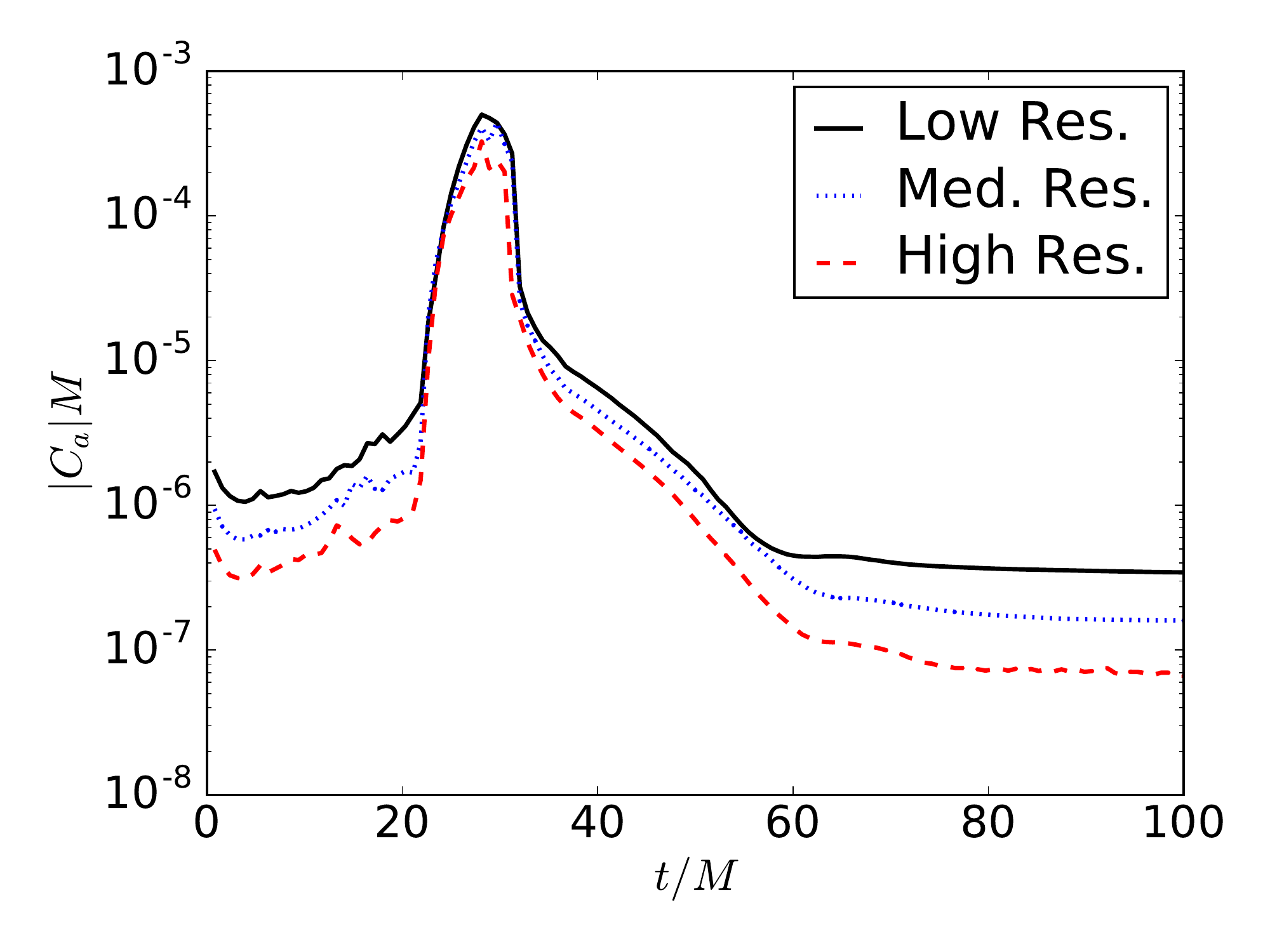}
\end{center}
\caption{
The L2 norm of the generalized
harmonic constraint violation $C_a=H_a-\Box x_a$ 
(average value in the $(x,y)\in [-20M,20M]\times[0,20M]$ central portion of the domain) 
at three resolutions for $b/M=10$ and $e=0.9$.
    In comparison to the low resolution, the grid resolution is $4/3\times$ and $2\times$ higher in the medium and high resolutions, respectively. The number of particles is $(4/3)^4\times$ and $2^4\times$ higher. 
\label{fig:cnst}
}
\end{figure}

In Fig.~\ref{fig:cnst}, we demonstrate the convergence of the Einstein
constraints for an example case with $b/M=10$ and $e=0.9$. During the initial and final stages of evolution,
the rate of convergence is close to second order, as expected 
(with the number of particles also increased to give this scaling).
Around the time the density blows up due to shell crossing (as described in the 
main text), the convergence is closer to first order, though this quickly improves
as an apparent horizon is found (the interior of which is excluded from this calculation).

Comparing the total energy in gravitational waves for the different
resolutions in this case, we estimate that the error in the dominant $\ell=2$ contribution is sub-percent
for the medium resolution.
The error in the very sub-dominant $\ell=4$ and 6 components (see Fig.~\ref{fig:egw}) is larger,
approximately $2\%$ and $30\%$, respectively.

\begin{figure}[H]
\begin{center}
\includegraphics[width=\columnwidth,draft=false]{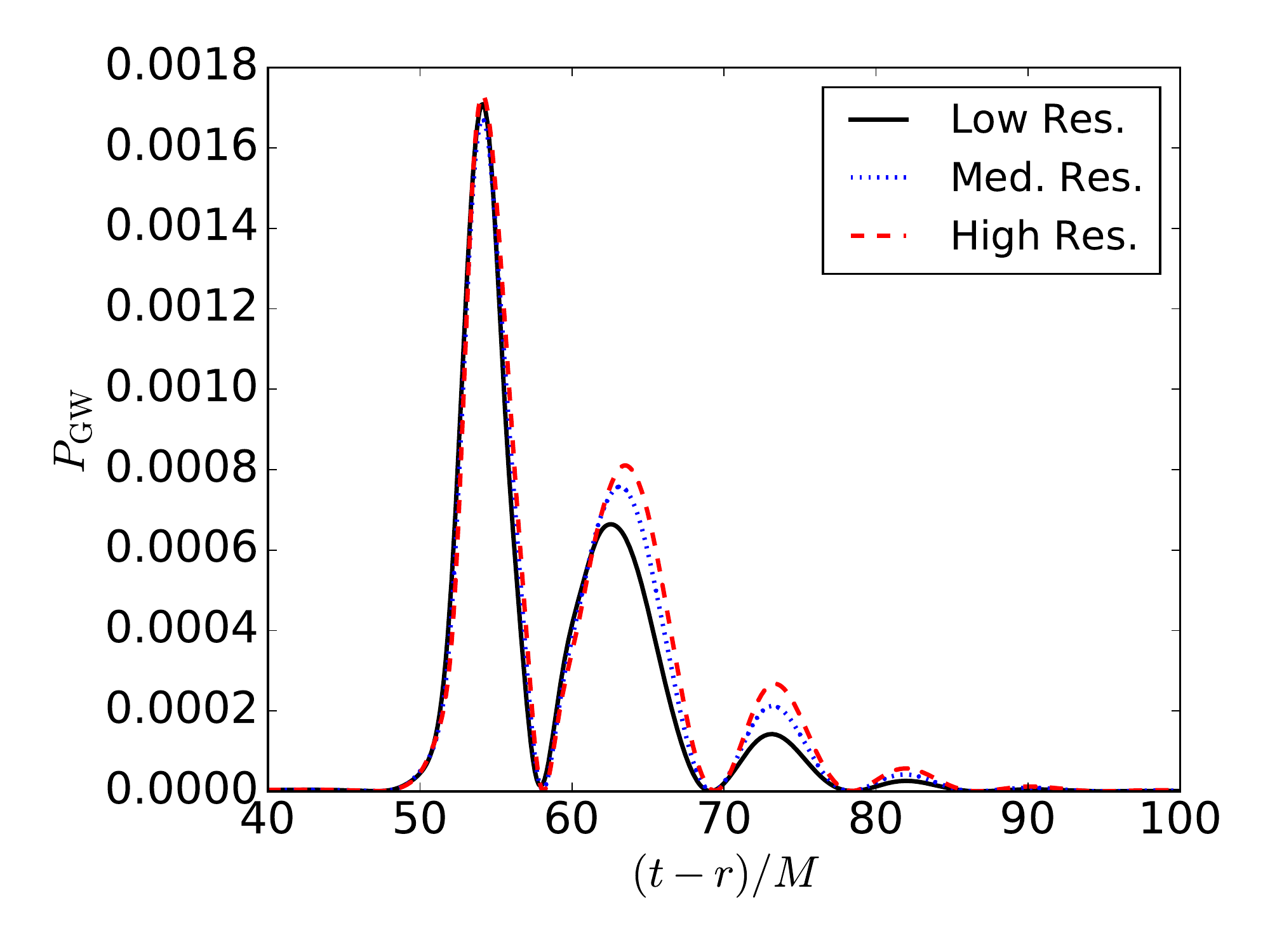}
\includegraphics[width=\columnwidth,draft=false]{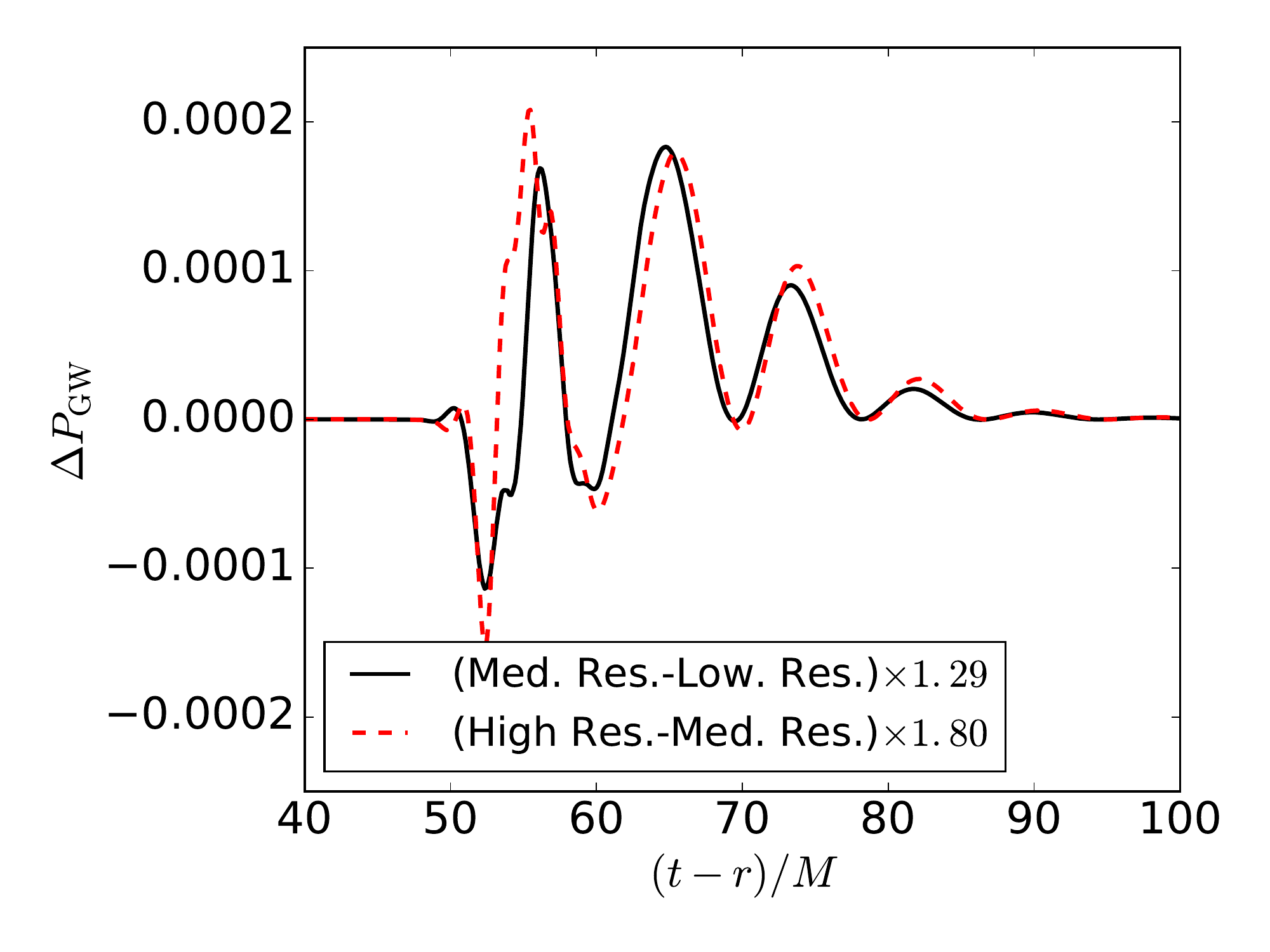}
\end{center}
\caption{
    The top panel shows the gravitational wave luminosity for the 
    $b/M=16$ and $e=0.9$ case at three resolutions.
    The bottom panel shows the difference in this quantity between resolutions, scaled assuming second order convergence.
	The relative grid resolution and number of particles is the same as in Fig.~\ref{fig:cnst}.
\label{fig:pgw_conv}
}
\end{figure}

We show the dependence of the gravitational wave luminosity
on resolution for the $b/M=16$ case in
Fig.~\ref{fig:pgw_conv}. The convergence in this
quantity is consistent with between first and second
order convergence. For this case, this error in the
total energy in gravitational waves is larger,
approximately $16\%$ for the medium resolution. 

\end{document}